\documentclass[pra,aps,twocolumn,nopacs,superscriptaddress,nofootinbib,longbibliography]{revtex4-1} %---APS,SI,arxiv with ref. titles %---use pra for SI
\usepackage{amsmath}  \usepackage{amssymb}  \usepackage{amsfonts}  \usepackage{bm}  \usepackage{bbm}  \usepackage{braket}  \usepackage{color}  \usepackage{comment}  \usepackage{dcolumn}  \usepackage{epsfig}
\usepackage{gensymb}  \usepackage{graphicx}  \usepackage{indentfirst}  \usepackage{lmodern}  \usepackage{mathrsfs}  \usepackage{mathtools}  \usepackage{psfrag}  \usepackage{pst-all}  \usepackage{soul}  \usepackage{units}  \usepackage{xcolor}
\usepackage{float} %---APS,SI,arxiv
\usepackage[colorlinks,linkcolor=blue,citecolor=blue,urlcolor=blue,hyperindex,driverfallback=dvipdfm]{hyperref}  \usepackage[T1]{fontenc} %---APS,ACS,SI,arxiv

\def\ii{{\rm i}}  \def\ee{{\rm e}}
  \def\kB{{k_{\rm B}}}
  \def\Imm{{\rm Im}}
          \def\eb{{\bf e}}                            \def\Qb{{\bf Q}}    \def\Rb{{\bf R}}  \def\rb{{\bf r}}    \def\ub{{\bf u}}  \def\vb{{\bf v}} %--- bold vectors
    \def\zz{\hat{\bf z}}            
\newcommand{\mb}[1]{\mathbf{#1}}

\begin{document} %---APS,arxiv
\def\bibsection{\section*{\refname}} %---SI,arxiv

% =========================================================
% --- title, affiliations, abstract -----------------------
% =========================================================
\title{Atomic-Scale Vibrational Mapping and Isotope Identification with Electron Beams
%\\{\color{gray} \small -- SUPPLEMENTARY INFORMATION -- } %---SI
}

% --- OSA affiliations ------------------------------------
%\author[1]{Andrea Kone\v{c}n\'{a}}
%\author[1]{
%\author{Fad{{\i}}l {{\.{I}}}y{{\i}}kanat}
%\author[1,2,*]{F.~Javier~Garc\'{\i}a~de~Abajo}
%\affil[1]{ICFO-Institut de Ciencies Fotoniques, The Barcelona Institute of Science and Technology, 08860 Castelldefels (Barcelona), Spain}
%\affil[2]{ICREA-Instituci\'o Catalana de Recerca i Estudis Avan\c{c}ats, Passeig Llu\'{\i}s Companys 23, 08010 Barcelona, Spain}
%\affil[*]{E-mail: javier.garciadeabajo@nanophotonics.es} %---

% --- APS,SI,arxiv affiliations ---------------------------
\author{Andrea Kone\v{c}n\'{a}}
\affiliation{ICFO-Institut de Ciencies Fotoniques, The Barcelona Institute of Science and Technology, 08860 Castelldefels (Barcelona), Spain}
\author{Fadil Iyikanat}
\affiliation{ICFO-Institut de Ciencies Fotoniques, The Barcelona Institute of Science and Technology, 08860 Castelldefels (Barcelona), Spain}
\author{F.~Javier~Garc\'{\i}a~de~Abajo}
\email{javier.garciadeabajo@nanophotonics.es}
\affiliation{ICFO-Institut de Ciencies Fotoniques, The Barcelona Institute of Science and Technology, 08860 Castelldefels (Barcelona), Spain}
\affiliation{ICREA-Instituci\'o Catalana de Recerca i Estudis Avan\c{c}ats, Passeig Llu\'{\i}s Companys 23, 08010 Barcelona, Spain}

% --- ACS affiliations ------------------------------------
%\author{Andrea Kone\v{c}n\'{a}}
%\affiliation[ICFO]{ICFO-Institut de Ciencies Fotoniques, The Barcelona Institute of Science and Technology, 08860 Castelldefels (Barcelona), Spain}
%\author{Fadil Iyikanat}
%\affiliation[ICFO]{ICFO-Institut de Ciencies Fotoniques, The Barcelona Institute of Science and Technology, 08860 Castelldefels (Barcelona), Spain}
%\author{F.~Javier~Garc\'{\i}a~de~Abajo}
%\email{javier.garciadeabajo@nanophotonics.es}
%\affiliation[ICFO]{ICFO-Institut de Ciencies Fotoniques, The Barcelona Institute of Science and Technology, 08860 Castelldefels (Barcelona), Spain}
%\alsoaffiliation[ICREA]{ICREA-Instituci\'o Catalana de Recerca i Estudis Avan\c{c}ats, Passeig Llu\'{\i}s Companys 23, 08010 Barcelona, Spain}

% --- document format -------------------------------------
%\begin{document} %---ACS
%\ociscodes{XXX} %(300.6530) Ultrafast spectroscopy; (270.1670) Coherent optical effects.} %---OSA, not SI

% --- abstract --------------------------------------------
\begin{abstract}
Transmission electron microscopy and spectroscopy currently enable the acquisition of spatially resolved spectral information from a specimen by focusing electron beams down to a sub-{\AA}ngstrom spot and then analyzing the energy of the inelastically scattered electrons with few-meV energy resolution. This technique has recently been used to experimentally resolve vibrational modes in 2D materials emerging at mid-infrared frequencies. Here, based on first-principles theory, we demonstrate the possibility of identifying single isotope atom impurities in a nanostructure through the trace that they leave in the spectral and spatial characteristics of the vibrational modes. Specifically, we examine a hexagonal boron nitride molecule as an example of application, in which the presence of a single isotope impurity is revealed through dramatic changes in the electron spectra, as well as in the space-, energy-, and momentum-resolved inelastic electron signal. We compare these results with conventional far-field spectroscopy, showing that electron beams offer superior spatial resolution combined with the ability to probe the complete set of vibrational modes, including those that are optically dark. Our study is relevant for the atomic-scale characterization of vibrational modes in novel materials, including a detailed mapping of isotope distributions.
\end{abstract}

% --- document format -------------------------------------
%\setboolean{displaycopyright}{true} %---OSA
%\begin{document} %---OSA
\maketitle %---APS,OSA,arxiv
\date{\today} %---APS,arxiv
\tableofcontents %---APS,SI,arxiv optional
%\setcounter{equation}{0} %---OSA
%\setkeys{acs}{maxauthors=0} %---ACS to avoid "et al." in references
%\noindent \textbf{Keywords:} electron energy-loss spectroscopy (EELS), isotope identification, vibrational spectroscopy, atomically resolved infrared spectroscopy, infrared microscopy %---ACS

% =========================================================
% --- introduction ----------------------------------------
% =========================================================
\section{Introduction}

The ability of exciting vibrational modes in crystals and molecules with localized probes has attracted much attention over the last decade because of the possibility of investigating chemical composition and atomic bonding with high spatial resolution. Numerous theoretical and experimental works have demonstrated that vibrational spectroscopy is feasible with nanoscale or even atomic-scale resolution using tip-based spectroscopic techniques, such as scanning near-field optical microscopy \cite{HTK02,HGA12,APN13} and tip-enhanced Raman spectroscopy \cite{KWK97,ZZD13,LCT19,JIM20}. These approaches rely on the electromagnetic optical-field enhancement produced at the probed sample area by introducing sharp metallic tips, such as those that are commonly used in atomic force and scanning tunneling microscopies. However, despite the substantial efforts made in engineering the geometric properties of the probing tip to improve resolution and sensitivity \cite{YZB06,AWH12,MGM18}, tip-based microscopies are still unable to map atomic vibrations, can only examine a fixed orientation of the specimen, and generally require complex analyses to subtract the undesired effects associated with tip-sample coupling. In addition, these techniques are only sensitive to vibrational modes that are optically or Raman active, while dark excitations without a net dipole moment remain difficult to detect.

Enabled by recent advances in instrumentation \cite{KUB09,KLD14}, electron energy-loss spectroscopy (EELS) performed in scanning transmission electron microscopes (STEMs) has emerged as an alternative, versatile technique capable of mapping vibrational modes with an atomic level of detail. In STEM-EELS, the sample is probed by a beam of fast electrons (typically with energies of 30 -- 300\,keV) focused below 1~\AA, thus allowing for the identification of individual atoms. Early theoretical predictions \cite{R14,D14,paper263,LK17,KNA18} were followed by experimental studies demonstrating the spectral and spatial characterization of low-energy excitations (10s -- 100s\,meV), such as phonons in general, phonon polaritons in nanostructured polar crystals \cite{KLD14,DAR16,LTH17,GKC17,ILT18,HNY18,KVM18,LB18,SSB19,QWL19}, molecular vibrations \cite{RAM16,HC18,JHH18,HHP19}, and hybrid modes resulting from the coupling between vibrational and plasmonic excitations \cite{paper342}. Recent achievements include the detection of vibrations at truly atomic resolution in hexagonal boron nitride (h-BN) \cite{HKR19}, silicon \cite{VLM19}, and graphene \cite{HRK20}, as well as the visualization of a single-atom impurity in the latter.

% Figure 1 ------------------------------------------------
\begin{figure*}
\centering
\includegraphics[width=1.00\textwidth]{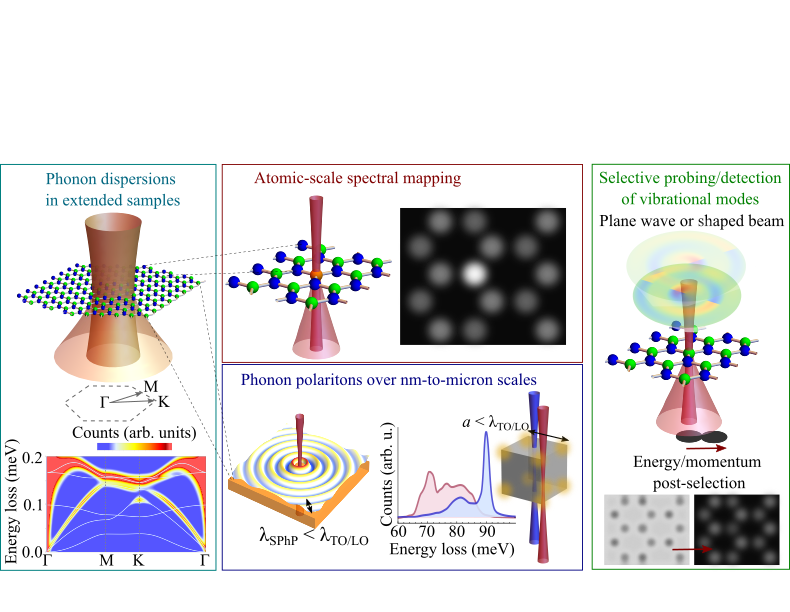}
\caption{Experimental approaches to probing vibrational excitations by EELS. Left: phonon dispersion in bulk materials can be studied through momentum-resolved EELS using an extended electron beam (e-beam) with a focal size of tens of nm \cite{HNY18,SSB19}. Middle top: localized vibrational modes in finite or defective structures can be probed at atomic resolution with tightly focused e-beams \cite{HKR19,VLM19,HRK20}. Middle bottom: phonon polaritons --electromagnetic waves coupled to the optical phonons of ionic crystals-- can efficiently be excited by e-beams to probe their spatial (nm-to-micron scale) and spectral (10s-to-100s meV) properties, which strongly depend on sample and probing geometry \cite{LTH17,GKC17,LK17}. Right: pre-shaping and post-selection of the electron transverse wave function enables the identification of specific vibrational-mode shapes and symmetries. Images are illustrations taken from our calculations.}
\label{Fig1} %{\red (adapted from ref\ \citenum{xxx})}
\end{figure*}

We depict common STEM-EELS experimental arrangements and types of probed infrared (IR) excitations in Figure~\ref{Fig1}. Dispersion relations of vibrational modes can be retrieved under broad electron beam (e-beam) irradiation of extended crystals combined with angle- and energy-resolved electron detection \cite{HNY18,SSB19} (left panel). In addition, localized vibrational modes (upper panel) can be probed with atomic detail using focused e-beams \cite{HKR19,VLM19,HRK20}, while the spatial and spectral distribution of phonon polaritons --hybrids of atomic vibrations and photons-- can also be mapped in structured samples \cite{LTH17,GKC17,LK17} (lower panel). Interestingly, by changing the collection conditions or the orientation of the specimen, one can filter different contributions to the inelastic signal \cite{paper243,D14,ZR20,RS21}, and similarly, selected excitations of specific symmetry may be addressed by shaping the electron wave function (right panel) in coordination with energy and momentum post-selection, as shown in a proof-of-principle experiment at higher excitation energies for the triggered detection of plasmons with dipolar and quadrupolar character \cite{GBL17}.

Here, we present first-principles calculations demonstrating the potential of e-beams for atomic-scale vibrational mapping, including the identification of single isotope-atom impurities. Specifically, we introduce a general computational methodology based on density-functional theory that can be applied to model spatially resolved EELS spectra at the atomic level. Such approach goes beyond the macroscopic dielectric formalism, which is successfully used to simulate low-loss EELS without atomic detail \cite{paper149,RTL17}, but fails to model the microscopic characteristics of the EELS signal. We show that e-beams are capable of exciting both bright and dark vibrational modes, while the latter are missed by far-field IR spectroscopy, as we illustrate by comparing our results to simulated optical absorption spectra. Additionally, we discuss polarization selectivity for different e-beam orientations with respect to the sample.

We note that atomically resolved vibrational mapping has so far been demonstrated only in samples that are resistant to e-beam damage, which is not the case of organic molecules, for which aloof EELS needs to be considered \cite{RAM16,HC18,JHH18,HHP19}. However, we foresee that less energetic electrons together with expected improvements in the sensitivity of electron analyzers may eventually allow us to probe molecular vibrations with high spatial resolution (see upper-mid panel in Figure~\ref{Fig1}). In preparation for these advances, we analyze here a rather stable h-BN-like molecule, which is purposely chosen as a model system that can naturally incorporate different boron isotopes (in particular $^{10}\rm{B}$ and $^{11}\rm{B}$) \cite{GDV18}, so it serves to study the effect of a single isotope defect on the resulting high-resolution EELS maps. In particular, we demonstrate that such impurity can significantly affect the vibrational modes and produce clearly discernible variations in the energy-filtered maps compared to those obtained for an isotopically pure sample. We thus conclude the feasibility of isotope identification at the single-atom level. In addition, we show that post-selection of scattered electrons depending on the acceptance angle of the spectrometer can improve these capabilities for isotope site recognition.

% =========================================================
% --- main text -------------------------------------------
% =========================================================
%\section{More sections} %---APS,OSA,SI,arxiv
%\section*{RESULTS AND DISCUSSION} %---ACS

% --- theory ----------------------------------------------
\section{Theoretical Description of Spatially Resolved Vibrational EELS}

For swift electrons of kinetic energy $>30\,$keV interacting with thin specimens (<10s\,nm) or under aloof conditions ({\it i.e.}, without actually traversing any material), coupling to each sampled mode is sufficiently weak as to be described within first-order perturbation theory, so that the EELS spectrum is contributed by electrons that have experienced single inelastic scattering events. As a safe assumption for e-beams, the initial ($i$) and final ($f$) electron wave functions can be separated as $\psi_{i|f}(\rb)=(1/\sqrt{L})\,\ee^{\ii p_{i|f,z} z}\psi_{i|f\perp}(\Rb)$, where $\Rb=(x,y)$ denotes the transverse coordinates, $L$ is the quantization length along the e-beam direction $z$, the longitudinal wave functions are plane waves of wave vectors $p_{{\rm i|f},z}$, and $\psi_{i|f\perp}(\Rb)$ are the transverse wave functions. In addition, we focus on scattering events that produce negligible changes in the electron momentum relative to the initial value, such that the electron velocity vector $\vb$ can be considered to remain constant (nonrecoil approximation). Finally, the atomic vibrations under study take place over small spatial extensions compared to the wavelength of light with the same frequency, so we adopt the quasistatic limit to describe the electron-sample interaction. Under these approximations, the EELS probability is given by \cite{paper149} (see Appendix)
\begin{widetext}
\begin{align}
\Gamma_{\rm EELS}(\omega)=\frac{e^2}{\pi\hbar v^2}\sum_{f}\int d^2\Rb\int d^2\Rb'\,&\psi_{i\perp}(\Rb)\psi^*_{f\perp}(\Rb)\psi^*_{i\perp}(\Rb')\psi_{f\perp}(\Rb') \label{GEELS}\\
&\times\int_{-\infty}^\infty dz\int_{-\infty}^\infty dz'\,\ee^{\ii\omega(z-z')/v}\,\Imm\left\{-W(\rb,\rb',\omega)\right\}, \nonumber
\end{align}
\end{widetext}
where we sum over final transverse states $f$ and the specimen enters through the screened interaction $W(\rb,\rb',\omega)$, defined as the potential created at $\rb$ by a unit charge placed at $\rb'$ and oscillating with frequency $\omega$. For atomic vibrations, the screened interaction reduces to a sum over the contributions of different vibrational modes, as shown in eqs\ \ref{WSS} and \ref{Snr} in the Appendix. Inserting these expressions into eq\ \ref{GEELS} and using the identity \cite{GR1980} $\int_{-\infty}^\infty dz\,\ee^{\ii qz}/r=2K_0(|q|R)$, where $K_0$ is a modified Bessel function, we find
\begin{align}
\Gamma_{\rm EELS}(\omega)=\frac{4e^2}{\pi\hbar v^2}\sum_{nf}\Imm\left\{
\frac{\left|N_{nfi}(\omega/v)\right|^2}{\omega_n^2-\omega(\omega+\ii\gamma)}
\right\}, \label{GEELSn}
\end{align}
where
\begin{align}
N_{nfi}(q)&=\int d^2\Rb\;\psi_{i\perp}(\Rb)\psi^*_{f\perp}(\Rb)I_n(\Rb,q), \nonumber\\
I_n(\Rb,q)&=\sum_l\frac{1}{\sqrt{M_l}}\,\int\,d^3\rb'\,K_0(|q||\Rb-\Rb'|)\,\ee^{\ii qz'}
\nonumber\\
&\quad\quad\quad\quad\quad\quad\quad\quad\quad\times \left[\eb_{nl}\cdot\vec{\rho}_l(\rb')\right], \label{InRq}
\end{align}
the $n$ sum runs over the sampled vibrational modes, $\omega_n$ and $\eb_{nl}$ are the associated frequencies and normalized atomic displacement vectors, respectively, the $l$ sum extends over the atoms in the structure, $M_l$ is the mass of atom $l$, the gradient of the charge distribution with respect to displacements of that atom is denoted $\vec{\rho}_l(\rb)$, and we have incorporated a phenomenological damping rate $\gamma$. As described in the Appendix, we use density-functional theory to obtain $\vec{\rho}_l(\rb')$ and the dynamical matrix. The latter is then used to find the natural vibration mode frequencies $\omega_n$ and eigenvectors $\eb_{nl}$ by solving the corresponding secular equation of motion. Because core electrons in the specimen are hardly affected by the swift electron, we assimilate them together with the nuclei to point charges $eZ_l$ located at the equilibrium atomic positions $\rb_l$, so that they contribute to $\vec{\rho}_l(\rb)$ with a term $\vec{\rho}_l^{\,\,\rm nucl}(\rb)=eZ_l\nabla_{\rb_l}\delta(\rb-\rb_l)$, which upon insertion into eq\ \ref{InRq} leads to
\begin{widetext}
\begin{align}
I_n(\Rb,q)=I_n^{\rm val}(\Rb,q)+\sum_l\frac{eZ_l}{\sqrt{M_l}}\,\ee^{\ii qz_l}\,\eb_{nl}\cdot\left[|q|\frac{(\Rb-\Rb_l)}{|\Rb-\Rb_l|}K_1(|q||\Rb-\Rb_l|)+\ii qK_1(|q||\Rb-\Rb_l|)\,\zz\right]. \nonumber
\end{align}
\end{widetext}
Here, $I_n^{\rm val}(\Rb,q)$ is computed from eq\ \ref{InRq} by substituting $\vec{\rho}_l(\rb)=\vec{\rho}_l^{\,\,\rm nucl}(\rb)+\vec{\rho}_l^{\,\,\rm val}(\rb)$ by the gradient of the valence-electron charge density $\vec{\rho}_l^{\,\,\rm val}(\rb)$ and carrying out the integral over a fine spatial grid. Incidentally, close encounters with the atoms in the structure produce an unphysical divergent contribution to the loss probability, which is avoided when accounting for the maximum possible momentum transfer \cite{paper149} and also when averaging over the finite width of the e-beam. For simplicity, we regularise this divergence in this work by making the substitution $|\Rb-\Rb'|\rightarrow\sqrt{|\Rb-\Rb'|^2+\Delta^2}$ with $\Delta=0.16\,${\AA} in the argument of the Bessel functions of the above equations.

Measurement of the inelastic electron signal in the Fourier plane corresponds to a selection of transmitted electron wave functions $\psi_{f\perp}(\Rb)=\ee^{\ii\Qb_f\cdot\Rb}/\sqrt{A}$ (normalized by using the transverse quantization area $A$), having a well-defined final transverse wave vector $\Qb_f\perp\zz$. Inserting this expression into eq\ \ref{GEELSn} and making the substitution $\sum_f\rightarrow[A/(2\pi)^2]\int d^2\Qb_f$, we find $\Gamma_{\rm EELS}(\omega)=\int d^2\Qb_f\,\left[d\Gamma_{\rm EELS}(\omega)/d\Qb_f\right]$, where
\begin{widetext}
\begin{align}
\frac{d\Gamma_{\rm EELS}(\omega)}{d\Qb_f}=\frac{e^2}{\pi^3\hbar v^2}\sum_n\left|\int d^2\Rb\;\psi_{i\perp}(\Rb)\ee^{-\ii\Qb_f\cdot\Rb}I_n(\Rb,\omega/v)\right|^2\Imm\left\{
\frac{1}{\omega_n^2-\omega(\omega+\ii\gamma)}
\right\} \label{EELSwQ}
\end{align}
\end{widetext}
is the momentum-resolved EELS probability. A dependence on the initial electron wave function is observed, which is however known to be lost when performing the integral over the entire $\Qb_f$ space \cite{RH1988,paper149}, leading to $\Gamma_{\rm EELS}(\omega)=\int d^2\Rb\;\left|\psi_{i\perp}(\Rb)\right|^2\Gamma_{\rm EELS}(\Rb,\omega)$, where
\begin{align}
\Gamma_{\rm EELS}(\Rb,\omega)=\frac{4e^2}{\pi\hbar v^2}\sum_n\Imm\left\{
\frac{\left|I_n(\Rb,\omega/v)\right|^2}{\omega_n^2-\omega(\omega+\ii\gamma)}
\right\} \label{EELSRw}
\end{align}
({\it i.e.}, the loss probability reduces to that experienced by a classical point electron averaged over the transverse e-beam density profile $\left|\psi_{i\perp}(\Rb)\right|^2$).

The above results are derived at zero temperature. When the vibrational modes are in thermal equilibrium at a finite temperate $T$, the loss probabilities given by eqs\ \ref{GEELSn}, \ref{EELSwQ}, and \ref{EELSRw} need to be corrected as $\Gamma_{\rm EELS}^T(\omega)=\Gamma_{\rm EELS}(|\omega|)\left[n_T(\omega)+1\right]{\rm sign}(\omega)$, where $n_T(\omega)=1/(\ee^{\hbar\omega/\kB T}-1)$ is the Bose-Einsten distribution function and $\omega>0$ ($\omega<0$) describes electron energy losses (gains). This expression, which has been used to determine phononic \cite{ILT18,LB18} and plasmonic \cite{paperarxiv5} temperatures with nanoscale precision using EELS, can be derived from first principles for bosonic modes \cite{paperarxiv2}, whose excitation and de-excitation probabilities are proportional to the quantum-harmonic-oscillator factors $n_T(\omega_n)+1$ and $n_T(\omega_n)$, respectively. In what follows, we ignore thermal corrections as a good approximation for energy losses $\hbar\omega\gtrsim50\,$meV at room temperature ($\kB T\sim\,25$meV).

% Figure 2 ------------------------------------------------
\begin{figure}
\centering
\includegraphics[width=0.45\textwidth]{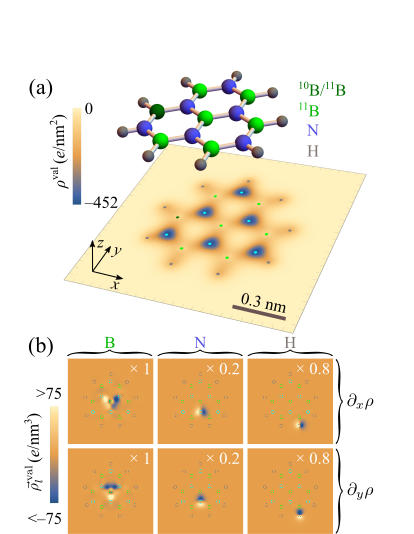}
\caption{Charge density gradients in a h-BN-like molecule. (a) We show the unperturbed valence-electron charge density integrated over the direction $z$ normal to the atomic plane for the molecular structure overlaid above the density plot. (b) Gradient of the $z$-integrated valence charge density with respect to in-plane atomic displacements along $x$ and $y$ for selected B, N, and H individual atoms (identified by the rapid variations of the gradient in these plots). The gradient is multiplied by a different factor in each plot (see labels) to maintain a common color scale.}
\label{Fig2}
\end{figure}

% --- results ---------------------------------------------
\section{Results and discussion}
\subsection{Valence-Electron Charge Gradients}

We focus on the model system sketched in Figure~\ref{Fig2}a, which consists of 7 boron (green) and 6 nitrogen (blue) atoms arranged in a h-BN-like configuration, and includes 9 hydrogen edge atoms (gray) to passivate the edges and give stability to the structure \cite{MO18} (see Appendix and supplementary Figure\ \ref{FigS6}). Molecules like this one are likely produced during chemical vapor deposition before a continuous h-BN film is formed \cite{MO18}, and they can also emerge when destroying continuous h-BN layers using physical methods \cite{VHW17}. In what follows, we compare the EELS signal from the isotopically pure molecule (taking all boron atoms as $^{11}\rm B$) with that obtained when one of the boron atoms is replaced by the isotope $^{10}\rm B$ (dark green ball in Figure~\ref{Fig2}a). The isotopic composition affects the vibrational modes, but not the valence-electron charge density, which is represented for the unperturbed molecule in the underlying color plot of Figure~\ref{Fig2}a (integrated over the $z$ direction, normal to the $x$-$y$ atomic plane). We observe electron accumulation around nitrogen atoms that reflects the ionic nature of the N-B bonds, similar to what is observed in extended h-BN monolayers. Coupling to the electron probe is mediated by the charge gradients $\vec{\rho}_l(\rb)$ entering eq\ \ref{InRq}, the valence contribution of which is shown in Figure~\ref{Fig2}b after integration over the out-of-plane direction ({\it i.e.}, $\int dz\,\vec{\rho}_l^{\,\,\rm val}(\rb)$) for the three types of atoms under consideration. Each atomic displacement leads to a dipole-like pattern centered around the displaced atom, and is qualitatively similar for other atoms of the same kind. As valence electrons pile up around nitrogen atoms, their displacements lead to the highest values of $\vec{\rho}_l^{\,\,\rm val}(\rb)$.

% Figure 3 ------------------------------------------------
\begin{figure*}
\centering
\includegraphics[width=\textwidth]{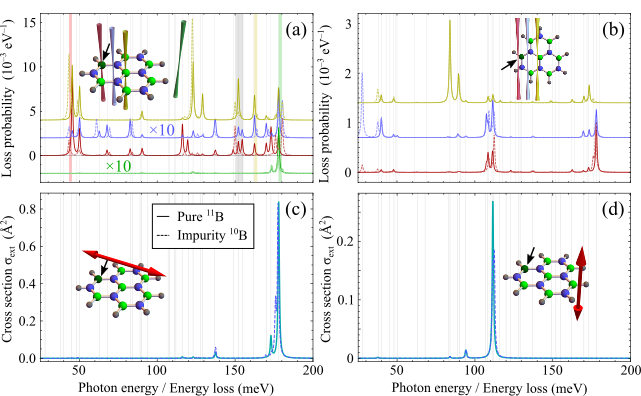}
\caption{Effect of an isotope impurity in EELS and infrared (IR) extinction. We compare calculated EELS (a,b) and optical extinction (c,d) spectra for an isotopically pure molecule (solid curves, only $^{11}$B atoms) and a molecule containing a single isotope atom impurity (dashed curves, $^{10}$B impurity marked by black arrows in the insets). For EELS (a,b), we take 60\,keV electrons and consider different e-beam positions (see color-coordinated vertically offset curves) and orientations (normal to the plane of the atoms in (a); parallel to and 1\,{\AA} away from that plane in (b)). For optical IR spectroscopy (c,d), we consider in-plane (c) and out-of-plane (d) light polarization (see red arrows in the insets). All vibrational mode energies obtained for the isotopically pure structure are indicated by vertical gray lines. We incorporate a spectral broadening of 1\,meV to account for intrinsic losses and instrument resolution.}
\label{Fig3}
\end{figure*}

% --- EELS vs IR spectroscopy -----------------------------
\subsection{Vibrational EELS {\it vs} Infrared Spectroscopy}

The gradients of the charge distribution in the nanostructure together with a vibrational eigenmode analysis provide the elements needed to calculate EELS and optical-extinction spectra in the IR range. In Figure~\ref{Fig3}a, we show EELS spectra obtained by using eq\ \ref{EELSRw} for 60\,keV electrons focused at four different positions within the $x$-$y$ atomic plane. We compare results for the isotopically pure molecule (solid curves) and the same structure with an isotopic impurity (dashed curves). The presence of the impurity leads to slight energy shifts of the mode energies ($\sim10$s\,meV, see details in supplementary Figures\ \ref{FigS1} and \ref{FigS2}), as well as substantial changes in the corresponding spectral weights. In addition, we observe that most features are enhanced when the e-beam is moved closer to the atomic positions, while several of them persist even when the beam passes outside the molecule ({\it cf.} blue and green spectra). Such persisting features are associated with modes that exhibit a net dipole moment, so they can also be revealed through far-field optical spectroscopy, as shown in the extinction cross sections plotted in Figure~\ref{Fig3}c (see Appendix for details of the calculation). Dipole-active modes couple to the e-beam over long distances, and consequently, they can be detected in the aloof configuration, which has been recently employed in EELS experiments to study beam-sensitive molecules \cite{RAM16,HC18,JHH18,HHP19}. Indeed, we corroborate a strong resemblance of the optical extinction (Figure~\ref{Fig3}c) and the aloof EELS (Figure~\ref{Fig3}a, green curves) spectra. Incidentally, the three-fold symmetry of the molecules renders the optical cross section independent of the orientation of the polarization vector within the plane of the atoms. The most intense EELS peaks are observed at around 130~meV and 170-180~meV, corresponding to modes involving B-N bond vibrations (see supplementary Figures\ \ref{FigS1} and \ref{FigS2} for details of the atomic displacement vectors associated with different vibrational modes). Weaker features at around 320~meV (see supplementary Figure\ \ref{FigS5}) arise from N-H bond stretches. Interestingly, the presence of the isotope impurity only has a small effect on dipole-active modes ({\it i.e.}, a small reduction of intensity and a weak energy shift). In contrast, some of the dark modes probed by EELS are severely affected.

By rotating the nanoflake with respect to the e-beam direction (or analogously, the light polarization for optical measuremensts), a different set of modes contributes to the spectral features, dominated by out-of-plane atomic motion under the conditions of Figure\ \ref{Fig3}b,d. In EELS, the strengths of these features strongly depend on e-beam position and orientation. In particular, the spectra shown in Figure\ \ref{Fig3}b (with the electrons passing parallel to and 1\,{\AA} away from the plane of the atoms) reveal low-energy peaks that are absent under normal incidence (Figure\ \ref{Fig3}a), associated with optical modes that are also missed in the optical extinction for out-of-plane polarization (Figure\ \ref{Fig3}d), which confirms their dark nature, in contrast to the excitations observed around $\sim100\,$meV energy.

% Figure 4 ------------------------------------------------
\begin{figure*}
\centering
\includegraphics[width=1.00\textwidth]{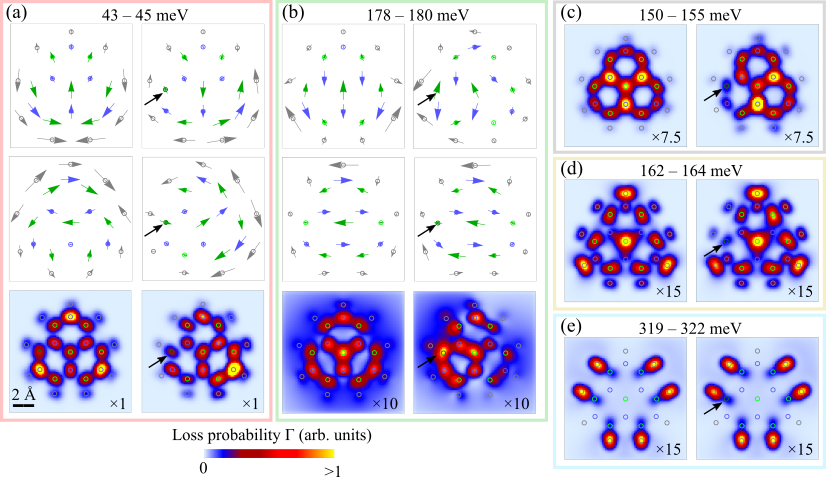}
\caption{Isotope-impurity determination through energy-filtered EELS imaging. We plot the calculated EELS probability integrated over different energy-loss ranges (see labels) as a function of e-beam position under the conditions of Figure\ \ref{Fig3}a for the isotopically pure molecule (left plots in each panel) and a molecule with an isotope atom defect (right plots, $^{10}$B atomic impurity indicated by black arrows). In (a,b), we show the atomic displacements associated with the involved vibrational modes. The probability is multiplied by a different factor in each density plot (see lower-right labels) to maintain a common color scale. Frame colors are coordinated with the vertical bands indicated in Figure\ \ref{Fig3}a. We spatially average the probability over a transverse Gaussian e-beam profile of 0.6\,{\AA} fwhm.}
\label{Fig4}
\end{figure*}

% --- atomic-scale mapping --------------------------------
\subsection{Vibrational Mapping at the Atomic Scale}

To visualize the complete spatial dependence of the vibrational EELS signal at specific energies corresponding to selected modes, we calculate energy-filtered maps by scanning the beam position over an area covering the studied structure (placed in the $x$-$y$ plane). In Figure\ \ref{Fig4}, we show maps calculated for a beam of 0.6\,{\AA} fwhm focal size and selected modes in isotopically pure and defective molecules (see supplementary Figures\ \ref{FigS3} and \ref{FigS4} for additional maps). By inspecting the results for $\sim44\,$meV (Figure\ \ref{Fig4}a) and $\sim179\,$meV (Figure\ \ref{Fig4}a) energy losses, together with the corresponding atomic displacement vectors (two degenerate modes at each of these energies), we observe a strong correlation in symmetry and strength between the mode displacements and the EELS maps, thus corroborating that the latter provide a solid basis to reconstruct the contribution of each atom to the vibrational modes. In addition, by introducing a boron isotope impurity (indicated by black arrows), the three-fold symmetry of the mode displacements and the resulting EELS maps are severely distorted. The impurity can produce either depletion or enhancement of the EELS intensity around its position, but in general, it influences the inelastic electron signal over the entire area of interest. However, for the B-H stretching modes around 320~meV, which are nearly decoupled from vibrations in different parts of the molecule, only the area close to the impurity is affected.

% Figure 5 ------------------------------------------------
\begin{figure*}
\centering
\includegraphics[width=1.00\textwidth]{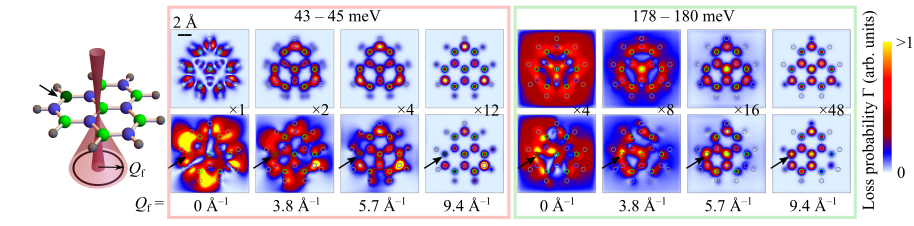}
\caption{Isotope-impurity determination through momentum-resolved EELS. Under the conditions of Figure\ \ref{Fig3}a, we plot the calculated EELS probability as a function of e-beam position for the isotopically pure molecule (upper plots) and a molecule with an isotope atom defect (lower plots, $^{10}$B atomic impurity indicated by black arrows) after integration over both the specified energy-loss range (around $\sim44\,$meV and $\sim179\,$meV, respectively) and the transverse wave vector of the transmitted electrons within a narrow annular aperture centered around the e-beam direction. Different radii of the latter ($Q_{\rm f}$) are considered in each column, as indicated by the lower labels. The probability is multiplied by a different factor in each color plot (see labels) to maintain a common color scale. The incident e-beam has a transverse Gaussian profile of 0.6\,{\AA} fwhm.}
\label{Fig5}
\end{figure*}

In Figure~\ref{Fig5}, we explore the effect of post-section on the resulting EELS intensity, as calculated for transverse-momentum-resolved scattered electrons using eq\ \ref{EELSwQ}. The e-beam is taken to have a Gaussian density profile $|\psi_i(\Rb)|^2$ of 0.6\,{\AA} fwhm. We present calculations for two of the most prominent features in EELS at energies around $\sim44\,$meV and $\sim179\,$meV, and compare results from isotopically pure (upper row) and defective (lower row) samples. The obtained momentum-dependent energy-filtered maps clearly demonstrate that the directly transmitted electrons ($Q_\mathrm{f}=0$) carry long-range information associated with the polarization of valence-electron charges (see supplementary Figures\ \ref{FigS3} and \ref{FigS4}, showing the separate contributions of valence-electron and nuclear charges). Such long-range signal is particularly prominent for the $\sim179\,$meV modes, which are dipole-active (see IR spectra in Figure~\ref{Fig3}b). In contrast, the modes around $\sim44\,$meV are dark, and thus, the map filtered at low momentum transfers reveals a weaker signal, except when symmetry is broken by the presence of the isotope impurity, which introduces a nonzero net dipole moment. In general, when collecting electrons experiencing larger transverse momentum transfers, the spatial localization of the signal is increased, enabling a better determination of the atomic positions and their contributions to the observed modes with a resolution limited by the finite size of the e-beam spot. Incidentally, we can link the energy-integrated low- and large-angle signal to bright- and dark-field images, respectively, as collected with different apertures in STEMs \cite{HKR19,ZR20}.

% =========================================================
% --- conclusion ------------------------------------------
% =========================================================
\section{Conclusions} %---APS,OSA,SI,arxiv
%\section*{CONCLUSIONS} %---ACS

STEM-EELS has evolved into a leading technique enabling both atomic-resolution imaging and spectroscopy over a broad frequency range extending down to the mid-IR. In particular, atomic-scale mapping of IR vibrational modes, which has recently become accessible \cite{HKR19,VLM19,HRK20}, is important for understanding and manipulating the optical response in such spectral range, as well as for determining the effect of phonons and phonon-polaritons in the electrical and thermal conductivities of nanostructured materials. We have shown that STEM-EELS can probe the complete set of vibrational modes, including optically bright and dark modes, the polarization characteristics of which can be addressed by tilting the sample relative to the e-beam or by resolving the inelastic electron signal in scattering angle. The latter can benefit from new advances in hybrid pixel detection technology \cite{PCD20}. As an application of these methods, we have demonstrated through first-principles calculations that an individual isotope impurity in a h-BN-like molecule produces radical changes in the spectral and spatial characteristics of the electron signal associated with the excitation of its vibrational modes. Our study supports the use of STEM-EELS to map isotope distributions with atomic precision, which can be important for understanding phononic lifetimes, as well as thermal and electrical transport at the nanoscale.

% =========================================================
% --- methods/appendix (not for SI) -----------------------
% =========================================================
\section*{APPENDIX} %---arxiv optional
%\section*{METHODS} %---ACS optional
%\begin{widetext} %---arxiv optional
%\appendix %---APS,OSA,arxiv optional
%\section*{Appendix} %---OSA optional
\renewcommand{\thesection}{A} %---OSA,arxiv optional
\renewcommand{\theequation}{A\arabic{equation}} %---OSA,arxiv optional
%\setcounter{equation}{0} %---OSA optional

% --- xxx -------------------------------------------------
%\noindent{\bf Title Format.} ... text ... %---ACS format for Methods section headers
%\section{title} ... text ... %---APS,OSA,arxiv format for Appendix section headers

% --- optical response ------------------------------------
\subsection*{EELS Probability} Under the assumptions discussed in the main text, the loss rate reduces to \cite{paper149}
\begin{align}
&\frac{\Gamma_{\rm EELS}(\omega)}{dt} \nonumber\\
&=\frac{2e^2}{\hbar}\sum_{f}\int d^3\rb\int d^3\rb'\,\psi_{i}(\rb)\psi^*_{f}(\rb)\psi^*_{i}(\rb')\psi_{f}(\rb') \nonumber\\
&\quad\quad\quad\quad\times\Imm\left\{-W(\rb,\rb',\omega)\right\}\,\delta(\varepsilon_f-\varepsilon_i+\omega), \nonumber
\end{align}
where $\psi_{i}(\rb)$ and $\psi_{f}(\rb)$ denote initial and final electron wave functions of energies $\hbar\varepsilon_i$ and $\hbar\varepsilon_f$, respectively. We now separate longitudinal and transverse components as specified in the main text ({\it i.e.}, $\psi_{i|f}(\rb))=(1/\sqrt{L})\,\ee^{\ii p_{i|f,z} z}\psi_{i|f\perp}(\Rb)$), use the nonrecoil approximation to write $\varepsilon_f-\varepsilon_i\approx(q_{f,z}-q_{i,z})v$, transform the sum over final states through the prescription $\sum_f\rightarrow(L/2\pi)\int dq_{f,z}\sum_f$ ({\it i.e.}, the remaining sum over $f$ now refers to transverse degrees of freedom), carry out the $q_{f,z}$ integral using the $\delta$ function, and multiply the result by the interaction time $L/v$ to convert the loss rate into a probability. Following these steps, we readily find eq\ \ref{GEELS}.

% --- optical response ------------------------------------
\subsection*{Optical Response Associated with Atomic Vibrations} The quasistatic limit is adopted here under the assumption that the studied structures are small compared with the light wavelength at the involved oscillation frequencies. We consider a perturbation potential $\phi^{\rm ext}(\rb,t)$ due to externally incident light or a swift electron, in response to which the atoms in the structure (labeled by $l=1,\cdots,N$) oscillate around their equilibrium positions $\rb_l$ with time-dependent displacements $\ub_l(t)$. The charge density $\rho({\{\ub\}},\rb)$, which obviously depends on $\{\ub\}\equiv\{\ub_1,\cdots,\ub_N\}$, is calculated from first principles as discussed below. Following a standard procedure to describe atomic vibrations \cite{AM1976}, we Taylor-expand the configuration energy (also calculated from first principles) for small displacements and only retain the lowest-order $\ub$-dependent contribution $(1/2)\sum_{ll'}\ub_l(t)\cdot\mathcal{D}_{ll'}\cdot\ub_{l'}(t)$, where $\mathcal{D}_{ll'}$ is the so-called dynamical matrix. We now write the Lagrangian of the system as $\mathcal{L}=(1/2)\sum_l M_l|\dot{\ub}_l(t)|^2-(1/2)\sum_{ll'}\ub_l(t)\cdot \mathcal{D}_{ll'}\cdot\ub_{l'}(t)-\int d^3\rb\,\rho({\{\ub\}},\rb)\phi^{\rm ext}(\rb,t)$, where $M_l$ is the mass of atom $l$ and the rightmost term accounts for the potential energy in the presence of the perturbing electric potential $\phi^{\rm ext}(\rb,t)$. The equation of motion then follows from $\partial_t\nabla_{\dot{\ub}_l}\mathcal{L}=\nabla_{\ub_l}\mathcal{L}$, which leads to
\begin{align}
M_l \ddot{\ub}_l(t)=-\sum_{l'}\mathcal{D}_{ll'}\cdot\ub_{l'}(t)-\int d^3\rb\,\vec{\rho}_l(\rb)\,\phi^\mathrm{ext}(\rb,t),
\label{Newton}
\end{align}
where we have used the property $\mathcal{D}^{\rm T}_{ll'}=\mathcal{D}_{l'l}$, while the vector field $\vec{\rho}_l(\rb)=\nabla_{\ub_l}\rho({\{\ub\}},\rb)$ represents the gradient of the electric charge density with respect to displacements of atom $l$. Linear response is assumed by using the dynamical matrix. In addition, to be consistent with this approximation, we evaluate $\vec{\rho}_l(\rb)$ at the equilibrium position $\ub=0$. It is then convenient to treat each frequency component separately to deal with the corresponding external potential $\phi^{\rm ext}(\rb,\omega)=\int_{-\infty}^\infty dt\,\ee^{\ii\omega t}\phi^{\rm ext}(\rb,t)$, so that eq\ \ref{Newton} becomes
\begin{align}
\omega(\omega+\ii\gamma)M_l\ub_l(\omega)=&\sum_{l'}\mathcal{D}_{ll'}\cdot\ub_{l'}(\omega) \nonumber\\
&+\int d^3\rb\,\vec{\rho}_l(\rb)\,\phi^\mathrm{ext}(\rb,\omega),
\label{Newton_exp}
\end{align}
where we have introduced a phenomenological damping rate $\gamma$. The solution to eq\ \ref{Newton_exp} can be found by first considering the symmetric eigenvalue problem for the free oscillations,
\begin{align}
\sum_{l'}\frac{1}{\sqrt{M_lM_{l'}}}\,\mathcal{D}_{ll'}\cdot\eb_{nl'}=\omega_n^2\,\eb_{nl},
\end{align}
where $n$ labels the resulting vibration modes of frequencies $\omega_n$. The eigenvectors $\eb_{nl}$ form a complete ($\sum_n \eb_{nl}^*\otimes\eb_{nl'}=\delta_{ll'}\mathcal{I}_3$, where $\mathcal{I}_3$ is the $3\times3$ unit matrix) and orthonormal ($\sum_l \eb_{nl}^*\cdot\eb_{n'l}=\delta_{nn'}$) basis set, which we use to write the atomic displacements as
\begin{align}
\ub_l(\omega)=\frac{1}{\sqrt{M_l}}\,\sum_nc_n(\omega)\,\eb_{nl}
\nonumber
\end{align}
with expansion coefficients
\begin{align}
c_n(\omega)=&\frac{1}{\omega(\omega+\ii\gamma)-\omega_n^2} \nonumber\\
&\times \sum_l\frac{1}{\sqrt{M_l}}\,\eb_{nl}^*\cdot\int d^3\rb\,\vec{\rho}_l(\rb)\,\phi^\mathrm{ext}(\rb,\omega).
\nonumber
\end{align}
Incidentally, because the dynamical matrix is real and symmetric, the eigenvectors $\eb_{nl}$ can be chosen to be real, but we consider a more general formulation using complex eigenvectors that are convenient to describe extended crystals, where the mode index $n$ may naturally incorporate a well-defined Bloch momentum. Now, introducing the above expression for $\ub_l$ in the induced charge density $\rho^{\rm ind}(\rb,\omega)=\sum_l\ub_l\cdot\vec{\rho}_l(\rb)$, we can obtain the susceptibility
\begin{align}
\chi(\rb,\rb',\omega)=\sum_{nll'}\frac{1}{\sqrt{M_lM_{l'}}}\,\frac{\left[\eb_{nl}\cdot\vec{\rho}_l(\rb)\right]\left[\eb_{nl'}^*\cdot\vec{\rho}_{l'}(\rb')\right]}{\omega(\omega+\ii\gamma)-\omega_n^2}, \label{chirrp}
\end{align}
which is implicitly defined by the relation $\rho^{\rm ind}(\rb,\omega)=\int d^3\rb'\,\chi(\rb,\rb',\omega)\,\phi^{\rm ext}(\rb',\omega)$. Finally, the screened interaction, defined by $W(\mb{r},\mb{r}',\omega)=\int d^3\rb_1\int d^3\rb_2\,\chi(\rb_1,\rb_2,\omega)\left|\rb-\rb_1\right|^{-1}\left|\rb'-\rb_2\right|^{-1}$, reduces to 
\begin{align}
W(\rb,\rb',\omega)=\sum_{n}\,\frac{S_{n}(\rb)S_{n}^*(\rb')}{\omega(\omega+\ii\gamma)-\omega_n^2}, \label{WSS}
\end{align}
where
\begin{align}
S_{n}(\rb)=\sum_l\frac{1}{\sqrt{M_l}}\int\,d^3\rb'\;\frac{\eb_{nl}\cdot\vec{\rho}_l(\rb')}{\left|\rb-\rb'\right|}. \label{Snr}
\end{align}
Equations\ \ref{WSS} and \ref{Snr} are used to evaluate eq\ \ref{GEELS} in the main text and produce eqs\ \ref{GEELSn}, \ref{EELSwQ}, and \ref{EELSRw}.

% --- extinction cross section ----------------------------
\subsection*{Optical Extinction Cross Section} For a small structure such as that considered in Figure\ \ref{Fig2}, we describe the far-field response in terms of the polarizability tensor, which we in turn calculate from the susceptibility by writing the dipole induced by a unit electric field, that is, $\bar{\bar{\alpha}}(\omega)=-\int d^3\rb\int d^3\rb'\,\chi(\rb,\rb',\omega)\,\rb\otimes\rb'$. Using eq\ \ref{chirrp}, we find
\begin{align}
\bar{\bar{\alpha}}(\omega)=\frac{2}{\hbar}\sum_n\frac{\omega_n\;{\bf d}_n\otimes{\bf d}_n^*}{\omega_n^2-\omega(\omega+\ii\gamma)}, \nonumber
\end{align}
where
\begin{align}
{\bf d}_n=\sum_{l}\sqrt{\frac{\hbar}{2\omega_nM_l}}\,\int d^3\rb\,\left[\eb_{nl}\cdot\vec{\rho}_l(\rb)\right]\;\rb \label{dn}
\end{align}
is the transition dipole associated with mode $n$. We then calculate the optical extinction cross section as \cite{V1981} $\sigma_{\rm ext}(\omega)=(4\pi\omega/c)\Imm\{\alpha_{ii}(\omega)\}$ for light polarization along either an in-plane ($i=x$) or an out-of-plane ($i=z$) symmetry direction. In practice, as described in the main text, we separate the charge gradient $\vec{\rho}_l(\rb)=\vec{\rho}_l^{\,\,\rm nucl}(\rb)+\vec{\rho}_l^{\,\,\rm val}(\rb)$ into the contributions of valence electrons ($\vec{\rho}_l^{\,\,\rm val}(\rb)$) and the rest of the system ($\vec{\rho}_l^{\,\,\rm nucl}(\rb)$, that is, nuclei and core electrons), so the transition dipoles reduce to ${\bf d}_n={\bf d}_n^{\rm val}+e\sum_{l}Z_l\sqrt{\hbar/(2\omega_nM_l)}\,\eb_{nl}$, where ${\bf d}_n^{\rm val}$ is calculated from eq\ \ref{dn} by replacing $\vec{\rho}_l(\rb)$ by $\vec{\rho}_l^{\,\,\rm val}(\rb)$.

% --- DFT -------------------------------------------------
\subsection*{DFT Calculations} We perform density functional theory (DFT) calculations using the projector-augmented-wave (PAW) method \cite{B94_2} as implemented in the Vienna \textit{ab initio} simulation package (VASP) \cite{KH93,KF96,KF96_2} within the generalized gradient approximation in the Perdew-Burke-Ernzerhof (PBE) form to describe electron exchange and correlation \cite{PBE96}. The cut-off energy for the plane waves is set to 500\,eV. The atomic positions of the molecular structures under consideration with and without an isotopic impurity are determined by minimizing total energies and atomic forces by means of the conjugate gradient method. Atomic positions are allowed to relax until the atomic forces are less than 0.02\,eV/{\AA} and the total energy difference between sequential steps in the iteration is below $10^{-5}$\,eV. The edges of the molecule are passivated with hydrogen terminations to maintain structural stability, leading to a nearly hexagonal structure (see supplementary Figure\ \ref{FigS6}). Additionally, hydrogen passivation eliminates a strong effect associated with the dangling bonds observed in the vibrational spectra. We calculate the dynamical matrix $\mathcal{D}_{ll'}$ using the small displacement method: each atom in the unit cell is initially displaced by 0.01\,{\AA} and the resulting interatomic force constants are then determined to fill the corresponding matrix elements. Vibrational eigenmodes and eigenfrequencies are obtained by diagonalizing the dynamical matrix, as discussed above. We assimilate nuclear and core-electron charges in each atom to a point charge. However, the distribution of the valence-electron charge density is incorporated using a dense grid in the unit cell to tabulate the charge gradients $\vec{\rho}_l^{\,\,\rm val}(\rb')$ from the change in the valence electron density produced for each small atomic displacement.

%\end{widetext} %---arxiv

% =========================================================
% --- SI, acknowledgments, bibliography, etc. -------------
% =========================================================

% --- SI --------------------------------------------------
%\section*{Supplementary Information} %---ACS optional
%The Supporting Information is available free of charge at https://pubs.acs.org/doi/xxx. Figures~S1 and S2: eigenenergies and eigenvectors for isotopically pure and defective structures. Figures~S3 and S4: nuclear+core-electron and valence-electron charge contributions to the energy-filtered EELS probability maps. Figure~S5: overview of spatially resolved EELS and IR absorption spectra covering a larger energy range. Figure~S6: relaxed atomic structures calculated with and without hydrogen edge passivation. %---ACS

% --- acknowledgments -------------------------------------
%\section*{Funding} % ... %---OSA for OSA, place funding here and thanks to colleagues below
\section*{Acknowledgments} %---ACS,APS,arxiv
%\begin{acknowledgments} %---APS
This work has been supported in part by the European Research Council (Advanced Grant 789104-eNANO), the European Commission (Horizon 2020 Grants 101017720 FET-Proactive EBEAM and 964591-SMART-electron), the Spanish MINECO (MAT2017-88492-R and Severo Ochoa CEX2019-000910-S), the Catalan CERCA Program, and Fundaci\'{o}s Cellex and Mir-Puig.
%\end{acknowledgments} %---APS

% --- bibliography (adapt file path as appropriate) -------
%\bibliographystyle{apsrev} %---APS,arxiv %---comment out for longbibliography
%\bibliography{../../../bibtex/refsL.bib} %---APS,OSA,arxiv with lower-case title format
%\bibliography{../../../bibtex/refsU.bib} %---ACS with upper-case title format

\clearpage %--- optional
\pagebreak \onecolumngrid \section*{Supplementary figures} %---SI,arxiv optional

% Figure S1 -----------------------------------------------
\begin{figure*}[h!]
\centering
\includegraphics[width=\textwidth]{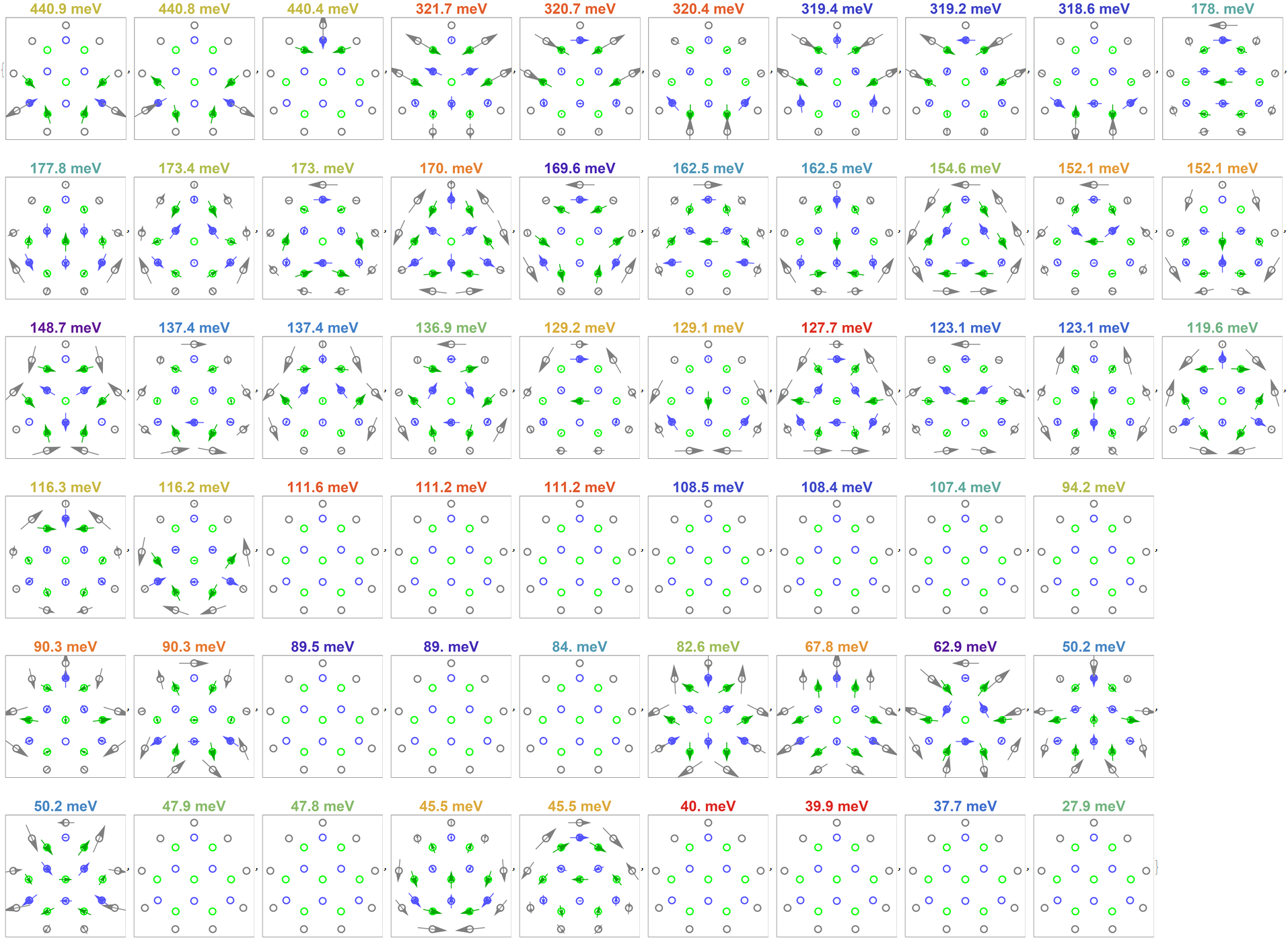}
\caption{Plots showing all of the eigenmodes and eigenvectors for the isotopically pure h-BN-like molecule in Figure\ 2a. Open circles denote the positions of B (green), N (blue), and H (gray) atoms, whereas the arrows show the magnitude and direction of the atomic discplacement vectors projected on the plane of the atoms. Mode energies are indicated above the plots, with (quasi-)degenerate ones sharing the same label color. We have excluded the 9 lowest energy modes, 6 of which emerge from rigid translations or rotations of the molecule.} 
\label{FigS1}
\end{figure*}

% Figure S2 -----------------------------------------------
\begin{figure*}[!h]
\centering
\includegraphics[width=\textwidth]{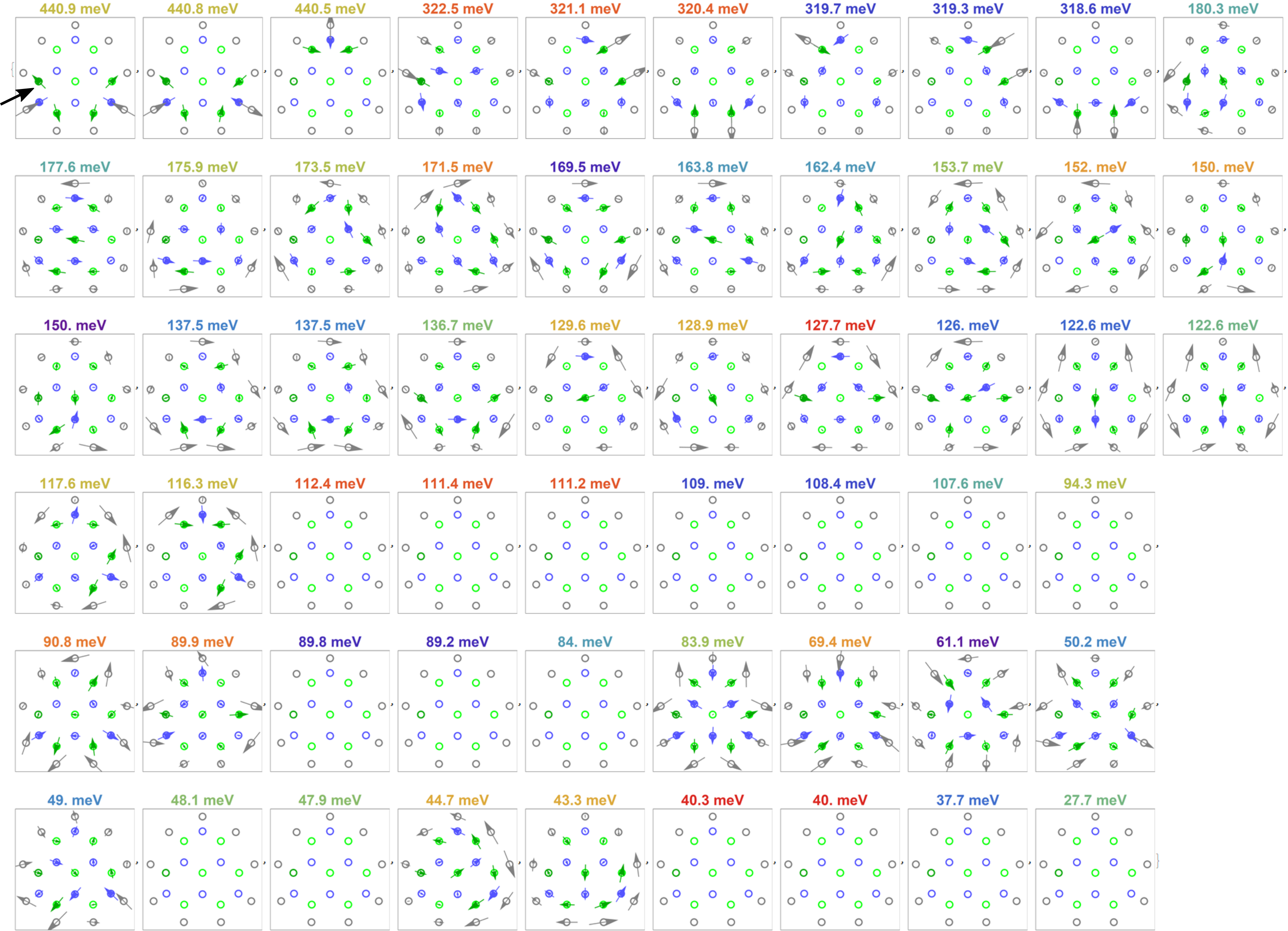}
\caption{Same as Fig.~\ref{FigS1}, but for the molecule containing a single B isotope impurity, represented by a dark green circles (see also black arrow in the upper-left panel and Figure\ 2a).} 
\label{FigS2}
\end{figure*}

% Figure S3 -----------------------------------------------
\begin{figure*}[!b]
\centering
\includegraphics[width=0.9\linewidth]{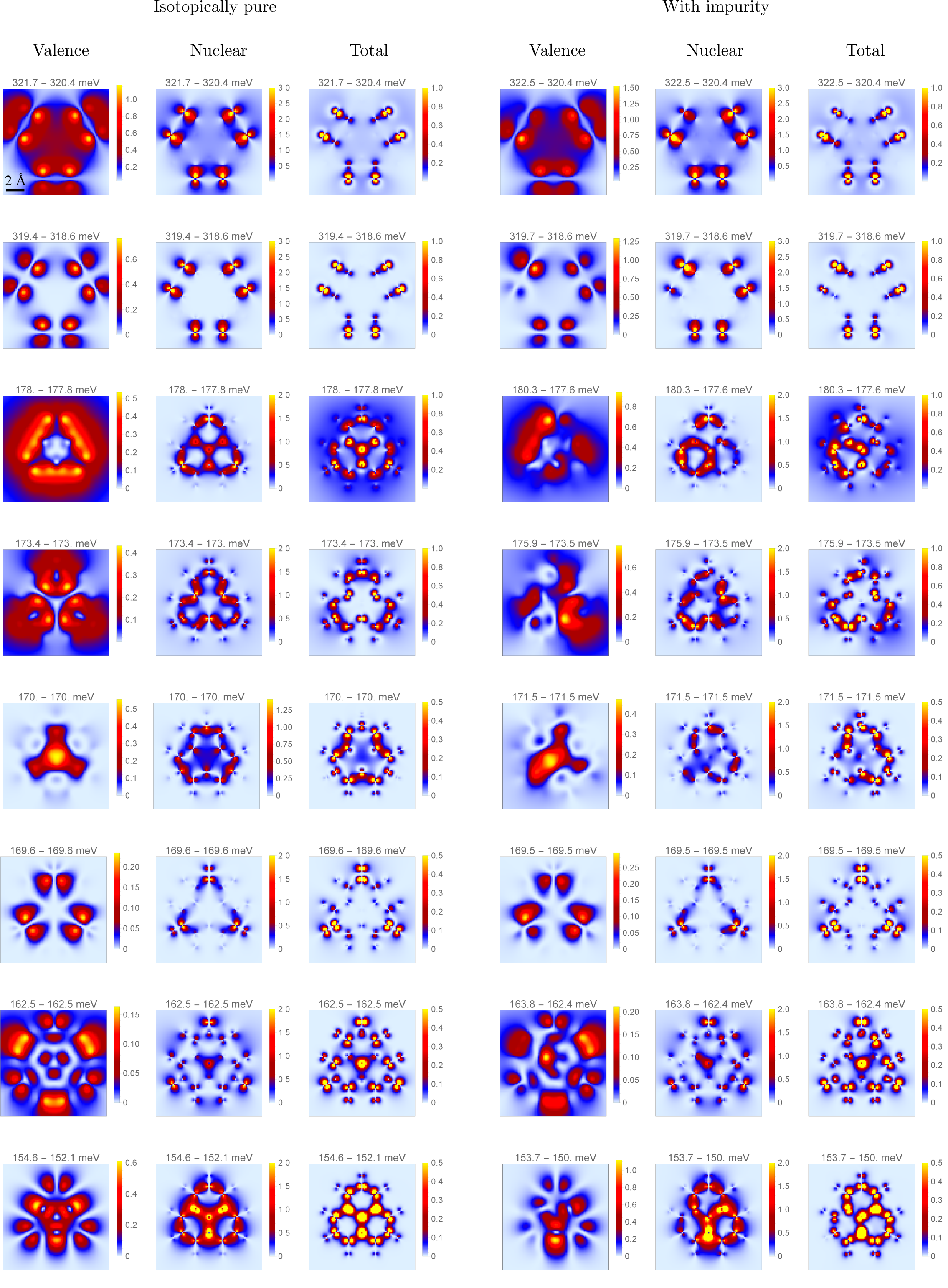}
\caption{Energy-filtered maps calculated as a function of the position of the focused electron beam by including different contributions to the polarization of the molecule associated with its vibrational modes (i.e., using the notation of the main paper, we consider partial contributions to $\vec{\rho}_l(\rb)=\vec{\rho}_l^{\,\,\rm nucl}(\rb)+\vec{\rho}_l^{\,\,\rm val}(\rb)$, where "nucl" refers to the sum of nuclear and core-electron charges, while "val" indicates the contribution of valence electrons). Different columns show calculations performed including valence-electron charges (valence), nuclear and core-electron charges (nuclear), and the sum of the two of them (total), as indicated by the upper labels. The three columns on the left (right) correspond to the isotopically pure (defective) molecule (see Figures\ \ref{FigS1} and \ref{FigS2}). Each map is obtained by integrating the EELS probability over the indicated energy range.}
\label{FigS3}
\end{figure*}

% Figure S4 -----------------------------------------------
\begin{figure*}[!b]
\centering
\includegraphics[width=0.9\linewidth]{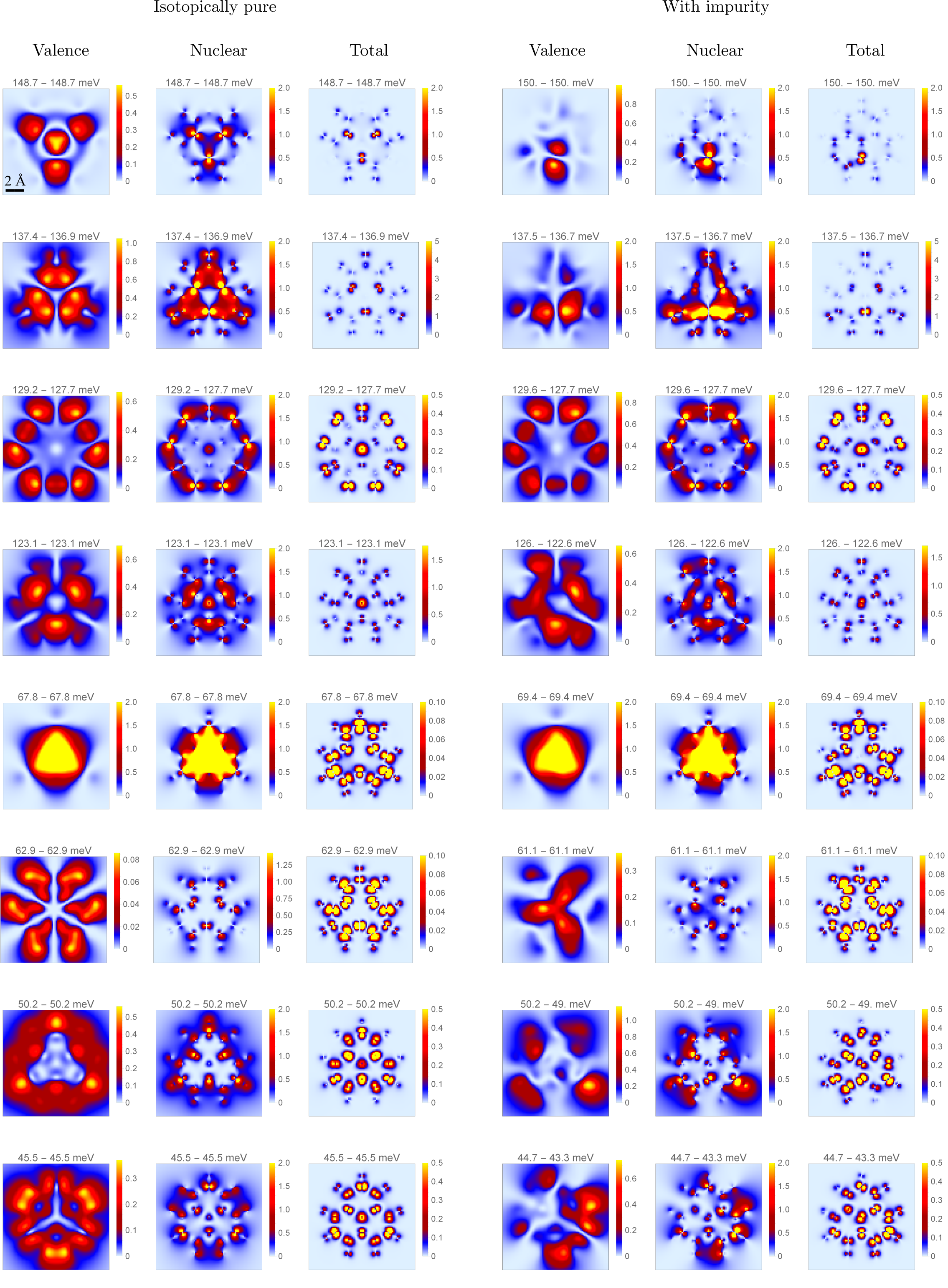}
\caption{Continuation of Figure\ \ref{FigS3}.}
\label{FigS4}
\end{figure*}

% Figure S5 -----------------------------------------------
\begin{figure*}[!ht]
\centering
\includegraphics[width=0.4\textwidth]{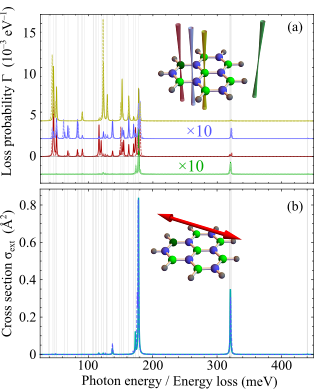}
\caption{Same as Figure\ 3a,c in the main text, but plotted over a wider energy range.} 
\label{FigS5}
\end{figure*}

% Figure S6 -----------------------------------------------
\begin{figure}[h]
\centering
\includegraphics[width=7cm]{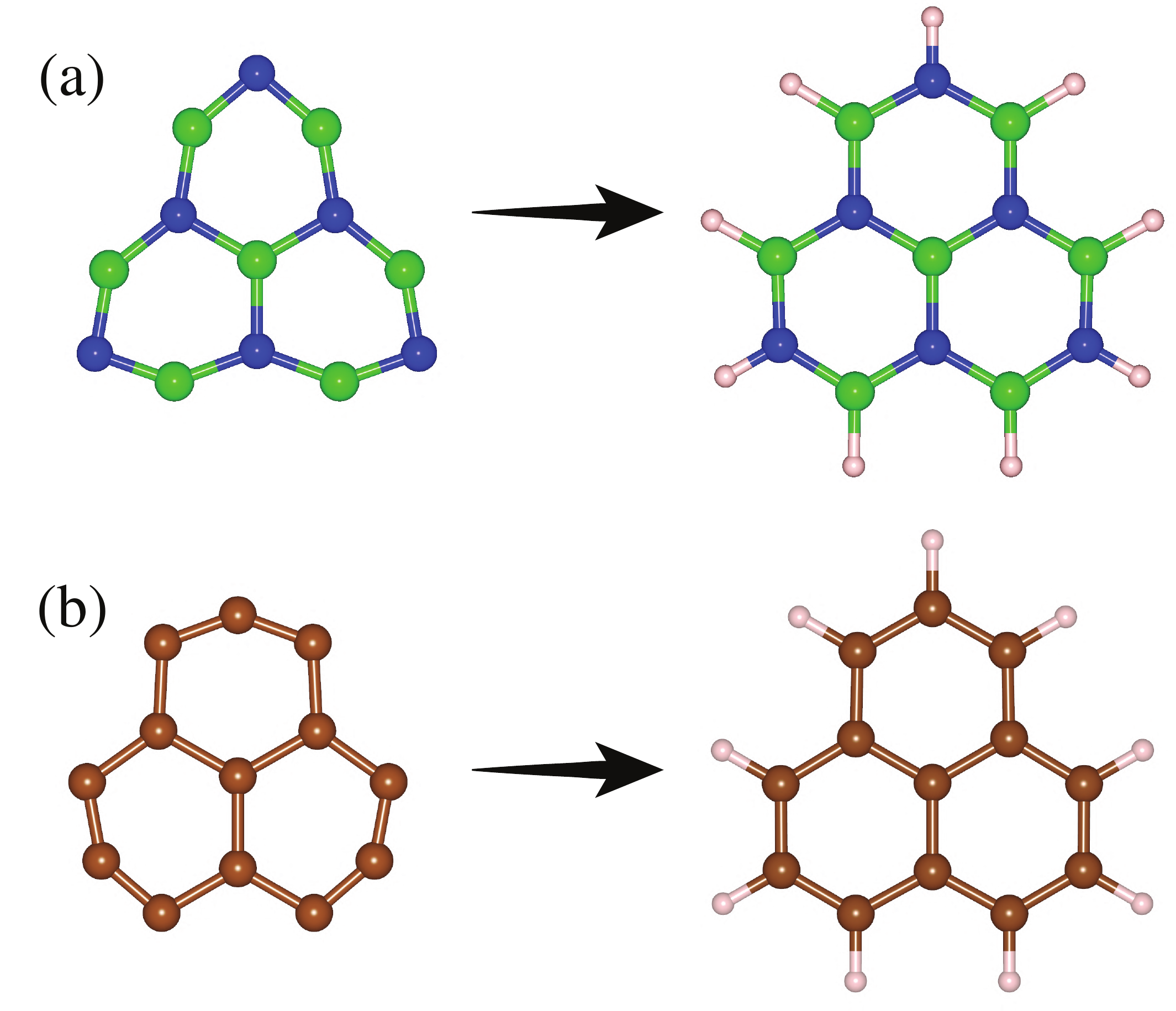}
\caption{Optimized atomic structure of (a) h-BN-like and (b) graphene molecules before (left) and after (right) passivation of the edges with additional hydrogen atoms.}
\label{FigS6}
\end{figure}

\end{document}